\documentclass[letterpaper, 10 pt, conference]{ieeeconf}
\IEEEoverridecommandlockouts                          
\usepackage{amsfonts}
\usepackage{amssymb}
\usepackage[cmex10]{amsmath}
\usepackage{graphicx}
\usepackage{subfigure}
\usepackage{verbatim}
\usepackage{cite}
\usepackage{fancyhdr}
\usepackage{tabularx}
\usepackage{booktabs}
\usepackage[ruled]{algorithm2e}
\usepackage{multirow}
\usepackage{hyperref}
\usepackage{bm}
\usepackage{stfloats}
\usepackage{enumerate}
\usepackage{wasysym}
\usepackage{amsmath}

\usepackage{amsthm}
\usepackage{fancyhdr}
\usepackage{amsfonts}
\usepackage{mathrsfs}
\usepackage{pifont}
\usepackage{amssymb}
\usepackage{verbatim}
\usepackage{upgreek}
\usepackage{graphicx}
\usepackage{epstopdf}
\usepackage{subfigure}
\usepackage{color}
\usepackage{algorithmic}
\newtheorem{Thm}{Theorem}
\newtheorem{Lem}{Lemma}
\newtheorem{Prop}{Proposition}
\newtheorem{Def}{Definition}

\newtheorem{Rmk}{Remark}

\title{\LARGE \bf
A Comparative Study of Artificial Potential Fields and Reciprocal Control Barrier Function-based  Safety Filters
}
\author{Ming Li$^{1}$ and Zhiyong Sun$^{2}$
\thanks{This work was supported in part by a research grant from National Science Foundation of China. \emph{(Corresponding author: Zhiyong~Sun.)}}
\thanks{$^{1}$M. Li is with the Division of Decision and Control Systems, KTH Royal Institute of Technology, Stockholm, Sweden ({\tt ming3@kth.se}).}
\thanks{$^{2}$Z. Sun is with Department of Mechanics and Engineering Science \& State Key Laboratory for Turbulence and Complex Systems, Peking University, Beijing, China ({\tt zhiyong.sun@pku.edu.cn}).}
}
\begin{document}

\maketitle
\thispagestyle{empty}
\pagestyle{empty}

\begin{abstract}
In this paper, we demonstrate that controllers designed by artificial potential fields (APFs) can be derived from reciprocal control barrier function quadratic program (RCBF-QP) safety filters. By integrating APFs within the RCBF-QP framework, we explicitly establish the relationship between these two approaches. Specifically, we first introduce the concepts of tightened control Lyapunov functions (T-CLFs) and tightened reciprocal control barrier functions (T-RCBFs), each of which incorporates a flexible auxiliary function. We then utilize an attractive potential field as a T-CLF to guide the nominal controller design, and a repulsive potential field as a T-RCBF to formulate an RCBF-QP safety filter. With appropriately chosen auxiliary functions, we show that controllers designed by APFs and those derived by RCBF-QP safety filters are equivalent. Based on this insight, we further generalize the APF-based controllers (equivalently, RCBF-QP safety filter-based controllers) to more general scenarios without restricting the choice of auxiliary functions. Finally, we present a collision avoidance example to clearly illustrate the connection and equivalence between the two methods.
\end{abstract}
\section{Introduction}
Safe autonomous motion planning of robotic systems is an increasingly important research topic, due to the growing presence of these systems in our daily lives.  Examples include self-driving cars navigating public roads \cite{claussmann2019review}, package delivery drones undergoing trials in urban environments \cite{murray2015flying}, and the coordination of humans and robots in warehouse \cite{wurman2008coordinating}. Among these applications, even a minor error in control can cause significant damage or jeopardize lives. As a result, ensuring safety (i.e., collision avoidance within this context) while allowing the robot to effectively pursue its navigation objectives is paramount. This requirement poses a significant challenge that continuously drives the evolution of motion planning techniques.

One of the classic methods in this domain is the~\textit{artificial potential fields} (APFs) approach, which was introduced over 30 years ago~\cite{APF_Initial} and is still widely used in mobile robot control. In the APF framework, target destinations are modeled as attractive potentials that pull the robot toward the goal, while obstacles are represented as repulsive potentials that push the robot away, thereby preventing collisions. This produces a control law that is both computationally efficient and intuitively interpretable~\cite{latombe1991robot}. However, despite its simplicity and ease of implementation, the standard APF method suffers from well-known issues such as getting trapped in local minima~\cite{bounini2017modified}, oscillatory behaviors near obstacles~\cite{ren2006modified}, and challenges in optimizing trajectory efficiency~\cite{krogh1984generalized}. To mitigate these problems, researchers have proposed various modifications to overcome local minima, and strategies for damping oscillations~\cite{ge2002dynamicAPF}.

In more recent years,~\textit{control barrier functions} (CBFs), which utilize Lyapunov-like arguments to ensure set invariance, have received significant attention~\cite{wieland2007constructive,ames2016CBF}.  They are often combined with control Lyapunov functions (CLFs) \cite{artstein1983stabilization,sontag1989Lyapunov}, in an attempt to find control inputs that guarantee both safety and stability. For example, CBF quadratic programs (CBF-QPs) modify a nominal controller (defined by CLFs), which prioritizes safety by making minimal adjustments to a nominal controller~\cite{ames2019control}. Extensions of this framework have addressed issues such as higher relative degree constraints~\cite{nguyen2016ECBF, xiao2021HOCBF}, robustness against disturbances and uncertainties~\cite{jankovic2018robust}, and decentralized control in multi-agent settings~\cite{jankovic2023multiagent}.

Regarding the above two methodologies, a comparative analysis was first conducted in~\cite{Comparative_Study} that attempted to address the question:  \textit{How do CBFs compare to APFs for obstacle avoidance tasks?} It has been demonstrated that ``APFs represent a particular case of CBFs: Given an APF, one can derive a CBF, while the reverse is not necessarily true." In this paper, we revisit this comparative study and aim to answer the following more fundamental questions: \textit{What is the relationship between the CBF-QP safety filter and the APF-designed controller?} Instead of using APFs to derive a zeroing-CBF (ZCBF) as in~\cite{Comparative_Study}, in this paper, we have demonstrated that APFs can be directly derived from reciprocal control barrier function quadratic program (RCBF-QP) safety filters.
 
The major differences between this paper and~\cite{Comparative_Study}, which are also the main contributions of this paper, are clarified as follows. In~\cite{Comparative_Study}, the authors first designed a ZCBF using an APF and then synthesized a controller using a~\textit{ZCBF-QP} safety filter. The resulting controller is characterized by a form~\textit{similar} to the one designed using APFs. In this paper, we bridge the gap between APFs and the~\textit{RCBF-QP} framework by explicitly demonstrating that these two approaches are~\textit{equivalent}. We have rigorously shown that APF-designed controllers can be derived from RCBF-QP safety filters. To achieve this, we introduce the concepts of tightened CLFs (T-CLFs) and tightened RCBFs (T-RCBFs), where each contains a flexible auxiliary function. We then utilize an attractive potential field as a T-CLF to guide the nominal controller design and a repulsive potential field as a T-RCBF to formulate an RCBF-QP safety filter. With appropriately chosen auxiliary functions, we demonstrate the equivalence between controllers designed through APFs and those derived from the RCBF-QP safety filter. Moreover, this insight allows us to generalize the APF-based controllers (equivalently, the RCBF-QP safety filter-based controllers) to more general scenarios without restricting the choice of auxiliary functions. Finally, we present a collision avoidance example to clearly illustrate the connection and equivalence between the two methods. 
\section{Preliminaries}
In this section, we provide a brief introduction to APFs, along with an overview of the concepts of CLFs, RCBFs, and RCBF-QP safety filters.
\subsection{Artificial Potential Fields} 
The fundamental idea of APFs is that obstacles within the environment repel the robot, while the goal position exerts an attractive force. Consequently, the robot navigates along the direction of the minimum potential energy, influenced by the combined effects of these forces. Although there are several variations of APF methods~\cite{modified_APF1,modified_APF2}, our particular focus lies in analyzing the original formulation in~\cite{APF_Initial}. Firstly, consider a control system described by a single-integrator dynamical model:
\begin{equation}\label{Single_Integrator}
    \dot{\mathbf{x}}=\mathbf{u},
\end{equation}
where $\mathbf{x}\in\mathbb{R}^{n}$ is the position, and $\mathbf{u}\in\mathbb{R}^{n}$ denotes the velocity, serving as the control input for the system. The objective is to formulate a desired velocity profile that steers the system towards a specified goal position while effectively avoiding obstacles encountered along the trajectory.
\subsubsection{Attractive Potential Field}
Firstly, we denote $\mathbf{x}_{\mathrm{goal}}$ as the goal position, which exerts an attractive potential field on the system, represented by:
\begin{equation}\label{Attractive}
    \mathrm{U}_{\mathrm{att}}(\mathbf{x})=\frac{1}{2}   K_{\mathrm{att}}\|\mathbf{x}-\mathbf{x}_{\mathrm{goal}}\|^{2},
\end{equation}
where $K_{\mathrm{att}}\in\mathbb{R }_{>0}$ is the attraction constant associated with the attraction potential field. The attraction force is obtained through the gradient of~\eqref{Attractive}, which is given as follows:
\begin{equation}\label{Attr_Force}
    \mathbf{F}_{\mathrm{att}}(\mathbf{x})=\nabla\mathrm{U}_{\mathrm{att}}(\mathbf{x})=K_{\mathrm{att}}\left(\mathbf{x}-\mathbf{x}_{\mathrm{goal}}\right),
\end{equation}
where $\nabla\mathrm{U}_{\mathrm{att}}(\mathbf{x})=\frac{\partial\mathrm{U}_{\mathrm{att}} }{\partial\mathbf{x}}(\mathbf{x})^{\top}$.
\subsubsection{Repulsive Potential Field}
Any obstacles in the area assert a repulsive potential field, given by
\begin{equation}\label{Repulsive}
\mathbf{U}_{\mathrm{rep}}(\mathbf{x})=\begin{cases}0, & \rho(\mathbf{x})\geq\rho_0, \\ \frac{1}{2} K_{\mathrm{rep}}\left(\frac{1}{\rho(\mathbf{x})}-\frac{1}{\rho_0}\right)^2 ,& \rho(\mathbf{x}) < \rho_0,\end{cases}
\end{equation}
where $K_{\mathrm{rep}}\in\mathbb{R}_{> 0}$ represents the repulsive constant associated with the repulsive potential field, and $\rho_0\in\mathbb{R}_{>0}$ is a user-defined parameter that determines the size of the region affected by repulsive potential field functions. Additionally, $\rho(\mathbf{x})$ denotes the distance to the obstacle or the distance from a safety area surrounding the obstacle, for instance:
\begin{equation}\label{Margin_function}
    \rho(\mathbf{x})=\|\mathbf{x}-\mathbf{x}_{\mathrm{obs}}\|-r,
\end{equation}
where  $\mathbf{x}_{\mathrm{obs}}$ is the position of the obstacle center, and $r\in\mathbb{R}_{>0}$ is the radius of the obstacle.  The repulsive force is obtained through the gradient of~\eqref{Repulsive}, which is given as follows:
\begin{equation}\label{Repulsive_Force}
\mathbf{F}_{\mathrm{rep}}(\mathbf{x})=\begin{cases}\mathbf{0}, & \rho(\mathbf{x})\geq\rho_0, \\ -\frac{K_{\mathrm{rep }}}{\rho(\mathbf{x})^2}\left(\frac{1}{\rho(\mathbf{x})}-\frac{1}{\rho_0}\right) \frac{\left(\mathbf{x}-\mathbf{x}_{\mathrm{obs }}\right)}{\|\mathbf{x}-\mathbf{x}_{\mathrm{obs }}\|}, & \rho(\mathbf{x}) < \rho_0.\end{cases}
\end{equation}
\subsubsection{Total Potential Field}
The attractive and repulsive potential fields are combined and the gradient is taken to obtain a feedback controller that pushes the robot to the goal while avoiding obstacles. Specifically, the total force resulting from the APF is the combination of the attractive force~\eqref{Attr_Force} and repulsive force~\eqref{Repulsive_Force}. 
\begin{equation}\label{APF_Force}
\mathbf{F}_{\mathrm{APF}}(\mathbf{x})=\begin{cases}
-\mathbf{F}_{\mathrm{att}}(\mathbf{x}), & \rho(\mathbf{x})\geq\rho_0, \\ -\mathbf{F}_{\mathrm{att}}(\mathbf{x})-\mathbf{F}_{\mathrm{rep}}(\mathbf{x}), & \rho(\mathbf{x}) < \rho_0.\end{cases}
\end{equation}
\subsubsection{APF Operational Mechanism}
\begin{figure}[tp]
 \centering
    \makebox[0pt]{%
    \includegraphics[width=0.4\textwidth]{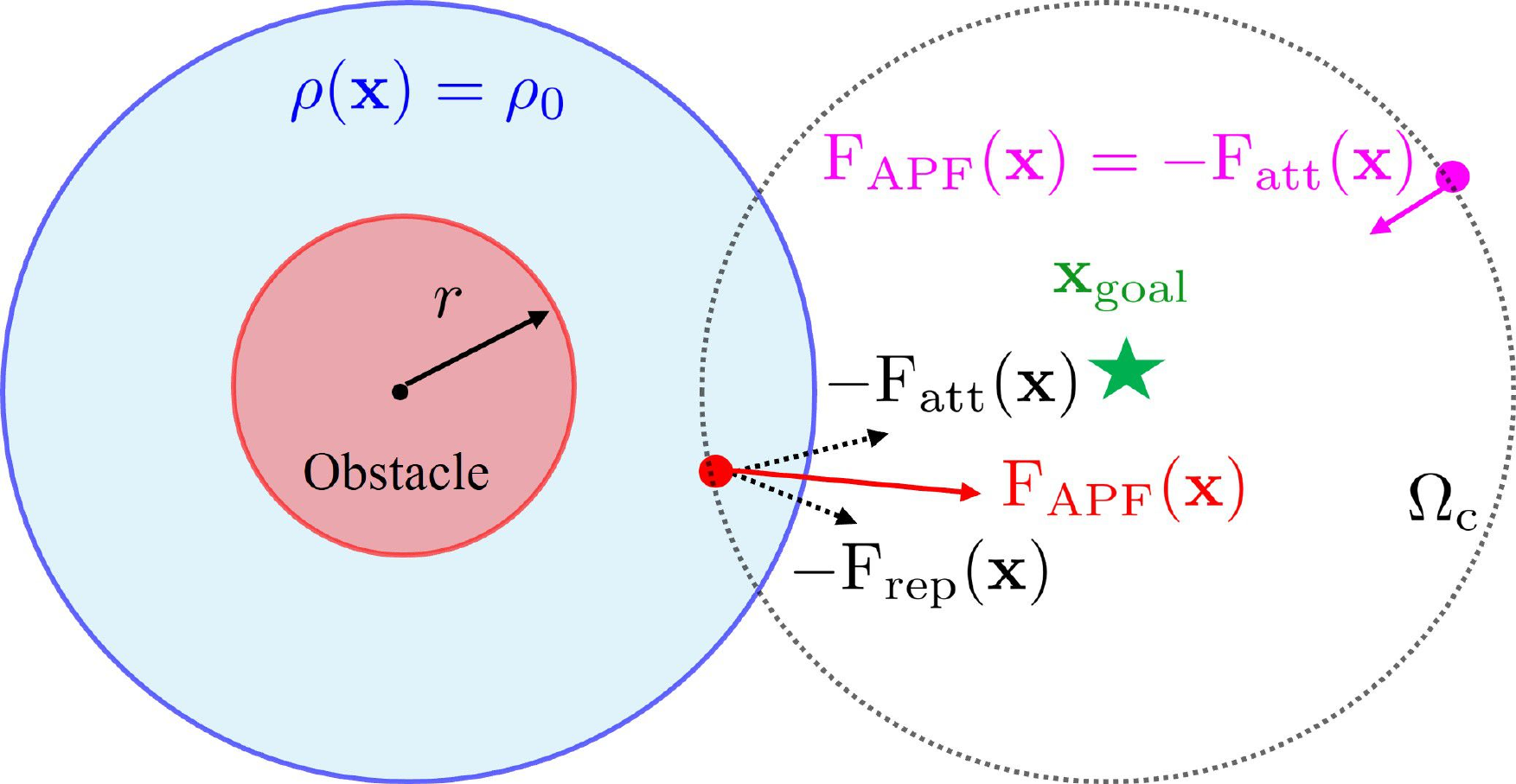}}
    \caption{APF Operational Mechanism: The shaded red area represents an obstacle with radius $r$, where the solid red circle denotes its boundary. The light blue shaded area illustrates regions where $\rho(\mathbf{x})<\rho_0$, with the solid blue circle indicating the boundary $\rho(\mathbf{x}) =\rho_0$. The green star denotes the goal position of a navigation task, while $\Omega_{\mathrm{c}}$ (black dashed circle) denotes the super level set of the attractive potential field $\mathrm{U}_{\mathrm{att}}(\mathbf{x})$. The red solid point signifies the robot, showcasing the force exerted on it when within the region where $\rho(\mathbf{x})<\rho_0$. Conversely, the magenta solid point depicts the attractive force exerted on the robot when it is situated within the region where $\rho(\mathbf{x})\geq \rho_0$.}.
    \label{APF}
\end{figure}
In Fig.~\ref{APF}, we provide a graphical interpretation of APF to explain its operational mechanism. For simplicity, we focus on a scenario featuring only one robot and one obstacle. As shown in Fig.~\ref{APF}, we define the super level set of the attractive potential field to be $\Omega_{\mathrm{c}}=\{\mathbf{x}\in\mathbb{R}^{n}|\mathrm{U}_{\mathrm{att}}(\mathbf{x})\leq\mathrm{c}\}$, where $\mathrm{c}\in\mathbb{R}_{>0}$. For the case that the robot (denoted by the red solid point) lies both on $\Omega_{\mathrm{c}}$ and in the light blue shaded area, i.e., $\rho(\mathbf{x}) < \rho_0$, the total force on the robot is $\mathbf{F}_{\mathrm{APF}}=-\mathbf{F}_{\mathrm{att}}(\mathbf{x})-\mathbf{F}_{\mathrm{rep}}(\mathbf{x})$ (red arrow), which is a combination of the force $-\mathbf{F}_{\mathrm{att}}(\mathbf{x})$ and $-\mathbf{F}_{\mathrm{rep}}(\mathbf{x})$ (associated with two dashed arrows in Fig.~\ref{APF}). When $\rho(\mathbf{x})\geq\rho_0$ (associated with the case that the robot is located on $\Omega_{\mathrm{c}}$ while outside the light blue shaded area), the control law is specified as $\mathbf{F}_{\mathrm{APF}}=-\mathbf{F}_{\mathrm{att}}(\mathbf{x})$ (depicted by the magenta arrow), wherein only attractive force is present.
\subsection{Control Lyapunov Functions}\label{CLF_CBF_QP}
Consider a control-affine dynamical system
	\begin{equation}\label{Affine_Control_System}
	    \dot{\mathbf{x}}=\mathbf{f}(\mathbf{x})+\mathbf{g}(\mathbf{x})\mathbf{u},
	\end{equation}
where $\mathbf{x}\in\mathbb{R}^n$ is the system state, $\mathbf{u}\in\mathbb{R}^m$ is the control input, $\mathbf{f}:\mathbb{R}^n\to\mathbb{R}^n$ is the drift vector field, and $\mathbf{g}:\mathbb{R}^n\to\mathbb{R}^{n\times m}$ is the control input matrix. It is assumed that both $\mathbf{f}$ and $\mathbf{g}$ are continuously differentiable functions, and $\mathbf{f}(\mathbf{0})=\mathbf{0}$. 
For system stabilization of the system~\eqref{Affine_Control_System}, CLF is a commonly-used tool, which is defined as follows.
\begin{Def}\label{CLF_Def}
(CLFs~\cite{sontag2013mathematical}) A continuously differentiable, positive definite, and radially unbounded function $V: \mathbb{R}^{n}\rightarrow\mathbb{R}_{\geq 0}$ is a CLF for the system \eqref{Affine_Control_System} if there exists a control input $\mathbf{u} \in \mathbb{R}^{m}$ satisfying
\begin{equation}\label{CLF_Condition}
\begin{aligned}
& a(\mathbf{x})+\mathbf{b}(\mathbf{x}) \mathbf{u}<0
\end{aligned}
\end{equation}
for all $\mathbf{x}\in\mathbb{R}^{n}\backslash \{\mathbf{0}\}$, where $a(\mathbf{x})=\frac{\partial V(\mathbf{x})}{\partial\mathbf{x}}\cdot\mathbf{f}(\mathbf{x})=L_{\mathbf{f}} V(\mathbf{x})$ and $\mathbf{b}(\mathbf{x})=\frac{\partial V(\mathbf{x})}{\partial\mathbf{x}}\cdot\mathbf{g}(\mathbf{x})=L_{\mathbf{g}} V(\mathbf{x})$. $L_{\mathbf{f}}$ and $L_{\mathbf{g}}$ denote the Lie derivatives along $\mathbf{f}$ and $\mathbf{g}$, respectively.
\end{Def}
\begin{Prop}
    (\cite[Proposition 5.9.10]{sontag2013mathematical}) Consider the nonlinear dynamical system~\eqref{Affine_Control_System} with a continuously differentiable, positive definite, and radially unbounded CLF $V: \mathbb{R}^{n}\rightarrow\mathbb{R}$. Then there must be a state feedback control law $\mathbf{u} = \mathbf{k}(\mathbf{x})$ with $\mathbf{k}(\mathbf{0}) = \mathbf{0}$ such that $a(\mathbf{x}) + \mathbf{b}(\mathbf{x}) \mathbf{k}(\mathbf{x}) < 0$ for all $\mathbf{x}\in\mathbb{R}^{n}\backslash \{\mathbf{0}\}$, which leads to the equilibrium point of \eqref{Affine_Control_System} being asymptoticly stable.
\end{Prop}

To illustrate the relationship between the APFs and the RCBF-QP safety filters (which will be discussed in Section~\ref{Compa_APF_CLF_CBF_QP}), a sufficient condition for the standard CLF condition~\eqref{CLF_Condition}, which is referred to the condition for defining a tightened CLF (T-CLF) in this paper, is introduced as follows:
\begin{equation}\label{Tightened_CLF_Condition}
    \mathop{\inf}_{\mathbf{u} \in \mathbb{R}^{m}}[\widetilde{a}(\mathbf{x})+\mathbf{b}(\mathbf{x})\mathbf{u}]\leq 0,
\end{equation}
where $\widetilde{a}(\mathbf{x})=a(\mathbf{x})+\sigma(\mathbf{x})$, and $\sigma(\mathbf{x})$ is a positive definite function. Typical choices of $\sigma(\mathbf{x})$ including $\sigma(\mathbf{x})=\kappa V(\mathbf{x})$ and $\sigma(\mathbf{x})=\zeta \|\mathbf{x}\|$, where $\kappa, \zeta\in\mathbb{R}_{>0}$ are both constants.  
\subsection{ Reciprocal Control Barrier Functions}
Unlike system stabilization which focuses on guiding a system state to a specific point or set, the notion of system safety is formalized by specifying a~\textit{safe set} in the state space that the system trajectory must remain in, i.e., not leaving a~\textit{safe set} if time evolves. In particular, we consider a closed set $\mathcal{C}\subset\mathbb{R}^{n}$ defined as the $0$-superlevel set of a continuously differentiable function $h:\mathbb{R}^{n}\rightarrow\mathbb{R}$, which is given by
\begin{equation}\label{Invariant_Set}
		\begin{aligned}
			\mathcal{C} & \triangleq\left\{\mathbf{x}\in \mathbb{R}^{n}: h(\mathbf{x}) \geq 0\right\}, \\
			\partial \mathcal{C} & \triangleq\left\{\mathbf{x}\in\mathbb{R}^{n}: h(\mathbf{x})=0\right\}, \\
			\operatorname{Int}(\mathcal{C}) & \triangleq\left\{\mathbf{x}\in \mathbb{R}^{n}: h(\mathbf{x})>0\right\},
		\end{aligned}
\end{equation}
where $\partial \mathcal{C}$ and $\operatorname{Int}(\mathcal{C})$ denote the boundary and interior of the set $\mathcal{C}$, respectively. 
\begin{Def}
    (Forward Invariance $\&$ Safety) A set $\mathcal{C}\subset\mathbb{R}^{n}$ is~\textit{forward invariant} if for every $\mathbf{x}_{0}\in\mathcal{C}$, the solution to~\eqref{Affine_Control_System} satisfies $\mathbf{x}(t)\in\mathcal{C}$ for all $t\in I(\mathbf{x}_{0})$. The system is~\textit{safe} on the set $\mathcal{C}$ if the set $\mathcal{C}$ is forward invariant. 
\end{Def}

As a ``dual'' concept of stability with CLFs, CBFs have been introduced to address safety-critical control problems~\cite{wieland2007constructive}. Generally, there are two main types of CBFs: zeroing CBFs (ZCBFs) and reciprocal CBFs (RCBFs). Although ZCBFs are more commonly used, this paper focuses on RCBFs because they are better suited for capturing repulsive potentials in APFs. Therefore, we provide the formal definition of RCBFs as follows.
\begin{Def}\label{RCBF_Def}
    (RCBFs~\cite{ames2016CBF}) Consider the control system~\eqref{Affine_Control_System} and the set $\mathcal{C}\subset\mathbb{R}^{n}$ defined by~\eqref{Invariant_Set} for a continuously differentiable function $h$. A continuously differentiable function $B:\mathrm{Int}(\mathcal{C})\rightarrow\mathbb{R}$ is called a RCBF if, for all $\mathbf{x}\in\mathrm{Int}(\mathcal{C})$, there exists class $\mathcal{K}$ functions $\alpha_{1},\alpha_{2},\alpha$ such that
\begin{subequations}\label{RCBF}
    \begin{align}
        \frac{1}{\alpha_1(h(\mathbf{x}))} \leq B(\mathbf{x}) &\leq \frac{1}{\alpha_2(h(\mathbf{x}))},\label{Zaa}\\
        \qquad\mathop{\inf}_{\mathbf{u} \in \mathbb{R}^{m}} [c(\mathbf{x})+\mathbf{d}&(\mathbf{x}) \mathbf{u}]\leq 0,\label{Zbb}
    \end{align}
\end{subequations}
where $c(\mathbf{x})=L_{\mathbf{f}}B(\mathbf{x})-\alpha(h(\mathbf{x}))$, $\mathbf{d}(\mathbf{x})=L_{\mathbf{g}} B(\mathbf{x})$.
\end{Def}

Similar to the tightened CLF condition as in~\eqref{Tightened_CLF_Condition}, we introduce a positive semidefinite function $\Gamma(\mathbf{x})$ into the condition~\eqref{Zbb} of the standard RCBF, which is given by 
\begin{equation}\label{Tightened_RCBF}
        \mathop{\inf}_{\mathbf{u} \in \mathbb{R}^{m}} [\widetilde{c}(\mathbf{x})+\mathbf{d}(\mathbf{x}) \mathbf{u}]\leq 0,
\end{equation}
where $\widetilde{c}(\mathbf{x})=c(\mathbf{x})+\Gamma(\mathbf{x})$, $\Gamma(\mathbf{x})$ is a positive semidefinite function. Note that we use the condition~\eqref{Zaa} and~\eqref{Tightened_RCBF} to define a tightened RCBF (T-RCBF). 

\subsection{RCBF-QP Safety Filters}
Firstly, we briefly introduce the RCBF-QP safety filter, which is defined by the following optimization problem:
\begin{subequations}\label{Tightened_CBF_QP}
    \begin{align}
         \mathbf{u}_{\mathrm{CBF}}=&\mathop{\mathrm{argmin}}_{\mathbf{u}\in\mathbb{R}^{m}}\frac{1}{2}\|\mathbf{u}-\mathbf{u}_{\mathrm{nom}}\|^{2}\label{Tightened_CBF_Cost}\\
    &\qquad\mathrm{s.t.}\,\, \widetilde{c}(\mathbf{x})+\mathbf{d}(\mathbf{x})\mathbf{u}\leq 0,\label{Tightened_CBF_Con}
    \end{align}
\end{subequations}
where $\mathbf{u}_{\mathrm{nom}}$ is a nominal controller. Note that in this formulation, we use the T-RCBF condition as the constraint in~\eqref{Tightened_CBF_QP}. This is because the latter choice of $\Gamma(\mathbf{x})$ provides flexibility to establish an equivalence between APFs and the RCBF-QP safety filter (which will be shown in the next section). Moreover, if we set $\Gamma(\mathbf{x})=0$, then constraint~\eqref{Tightened_CBF_Con} immediately reduces to RCBF condition~\eqref{Zbb}.
\begin{Lem}\label{Tightened_CBF_QP_Lem}
    The explicit solution to the RCBF-QP safety filter, i.e., the optimization problem~\eqref{Tightened_CBF_QP}, is:
    \begin{equation}\label{Safety_Filter}
        \mathbf{u}_{\mathrm{CBF}}=\begin{cases}
        \mathbf{u}_{\mathrm{nom}},&\varphi(\mathbf{x})\leq 0,\\
        \mathbf{u}_{\mathrm{nom}}-\frac{\varphi(\mathbf{x})}{\|\mathbf{d}(\mathbf{x})\|^{2}}\mathbf{d}(\mathbf{x})^{\top},&\varphi(\mathbf{x})> 0,
\end{cases}
    \end{equation}
    where $\varphi(\mathbf{x})=\widetilde{c}(\mathbf{x})+\mathbf{d}(\mathbf{x})\mathbf{u}_{\mathrm{nom}}$.
\end{Lem}
\vspace{3pt}
\begin{proof}
See~\cite{cohen2024safety}.
\end{proof}
\section{APFs versus RCBF-QP Safety Filters}\label{Compa_APF_CLF_CBF_QP}
In this section, our goal is to bridge the gap between the APFs and the RCBF-QP safety filters, demonstrating that the controllers designed using APFs are equivalent to those of the RCBF-QP safety filters.
\subsection{Asymptotic Stability and Safety Guarantees with APFs}
Firstly, we use the attractive potential field and the repulsive potential field from~\eqref{Attractive} and~\eqref{Repulsive} as the CLF and RCBF, respectively, for the single-integrator model in~\eqref{Single_Integrator}. We then show that the corresponding control laws, defined by the attractive force~\eqref{Attr_Force} and the repulsive force~\eqref{Repulsive_Force}, guarantee asymptotic stability and safety of the system~\eqref{Single_Integrator}, respectively.
\begin{Lem}\label{CLF_Att}
  Consider the single-integrator model~\eqref{Single_Integrator} and set $V(\mathbf{x})=\mathrm{U}_{\mathrm{att}}(\mathbf{x})$, where $\mathrm{U}_{\mathrm{att}}(\mathbf{x})$ is the attractive potential field defined in~\eqref{Attractive}. The control law $\mathbf{u}=-\mathbf{F}_{\mathrm{att}}(\mathbf{x})$, defined by the attractive force given in~\eqref{Attr_Force}, satisfies the CLF condition~\eqref{CLF_Condition} and hence ensures the asymptotic stability of the system~\eqref{Single_Integrator}.
\end{Lem}
\begin{proof}
Firstly, by setting $V(\mathbf{x})=\mathrm{U}_{\mathrm{att}}(\mathbf{x})$ for the dynamical model~\eqref{Single_Integrator}, we obtain $a(\mathbf{x})=0$ and $\mathbf{b}(\mathbf{x})=\mathbf{F}_{\mathrm{att}}(\mathbf{x})^{\top}$ with Definition~\ref{CLF_Def}. Moreover, according to the definition of attractive force in~\eqref{Attr_Force}, we know that $\|\mathbf{F}_{\mathrm{att}}(\mathbf{x})\|^{2}>0$ for all $\mathbf{x}\in\mathbb{R}\backslash\{\mathbf{x}_{\mathrm{goal}}\}$. Next, we substitute the control law $\mathbf{u}=-\mathbf{F}_{\mathrm{att}}(\mathbf{x})$ into the CLF condition~\eqref{CLF_Condition}, which gives
\begin{equation}
    -\mathbf{F}_{\mathrm{att}}(\mathbf{x})^{\top}\mathbf{F}_{\mathrm{att}}(\mathbf{x})< 0,\quad\forall\mathbf{x}\in\mathbb{R}\backslash\{\mathbf{x}_{\mathrm{goal}}\}.
\end{equation}
Then we conclude that the closed-loop system~\eqref{Single_Integrator} is asymptotically stable.
\end{proof}
\begin{Lem}\label{CBF_Rep}
  Consider the single-integrator model~\eqref{Single_Integrator} and set $B(\mathbf{x})=\mathbf{U}_{\mathrm{rep}}(\mathbf{x})$ and $h(\mathbf{x})=\rho(\mathbf{x})$, where the repulsive potential field $\mathbf{U}_{\mathrm{rep}}(\mathbf{x})$ and the function $\rho(\mathbf{x})$ are given in~\eqref{Repulsive} and~\eqref{Margin_function}, respectively. The control law $\mathbf{u}=-\mathbf{F}_{\mathrm{rep}}(\mathbf{x})$, defined by the repulsive force given in~\eqref{Repulsive_Force}, satisfies the RCBF condition and thus ensures the safety of the system~\eqref{Single_Integrator}.
\end{Lem}
\begin{proof}
To prove that the repulsive potential field $B(\mathbf{x})=\mathbf{U}_{\mathrm{rep}}(\mathbf{x})$ with  $\mathbf{u}=-\mathbf{F}_{\mathrm{rep}}(\mathbf{x})$ satisfies the RCBF condition, we need to verify that i) $B(\mathbf{x})$ satisfies the condition~\eqref{Zaa}, and ii) the condition~\eqref{Zbb} holds.

Regarding the first aspect, as depicted in Fig.~\ref{APF}, the system remains safe when $\rho(\mathbf{x}) \geq \rho_0$. Then, we only need to examine the situation where $0\leq \rho(\mathbf{x})< \rho_0$. In this case, the function $B(\mathbf{x})$ can be formulated as follows:
\begin{equation}
    B(\mathbf{x})=\mathbf{U}_{\mathrm{rep}}(\mathbf{x})=\frac{1}{\overline{\alpha}(h(\mathbf{x}))}, 
\end{equation}
where 
\begin{equation}
    \overline{\alpha}(h(\mathbf{x}))=\frac{2}{K_{\mathrm{rep}}}\cdot\left(\frac{\rho_0 h(\mathbf{x})}{\rho_0-h(\mathbf{x})}\right)^{2}.
\end{equation}
Here, it can be verified that $\overline{\alpha}(h(\mathbf{x}))$ is a class $\mathcal{K}$ function when $0\leq \rho(\mathbf{x})< \rho_0$ (note that $h(\mathbf{x})=\rho(\mathbf{x})$), and hence we can always determine class functions $\alpha_{1}(h(\mathbf{x}))$ and $\alpha_{2}(h(\mathbf{x}))$ to ensure that the condition~\eqref{Zaa} is satisfied.

Furthermore, when $0\leq \rho(\mathbf{x})< \rho_0$, according to~\eqref{Repulsive_Force} and~\eqref{RCBF}, we have $c(\mathbf{x})=-\alpha(h(\mathbf{x}))$ and $\mathbf{d}(\mathbf{x})=\mathbf{F}_{\mathrm{rep}}(\mathbf{x})^{\top}$ for the dynamical model~\eqref{Single_Integrator}.  By substituting $\mathbf{u}=-\mathbf{F}_{\mathrm{rep}}(\mathbf{x})$ into~\eqref{Zbb}, it yields
    \begin{equation}\label{Safety_Condition}
    -\alpha(h(\mathbf{x}))-\mathbf{F}_{\mathrm{rep}}(\mathbf{x})^{\top}\mathbf{F}_{\mathrm{rep}}(\mathbf{x})< 0
\end{equation}
since $\|\mathbf{F}_{\mathrm{rep}}(\mathbf{x})\|^{2}>0$ for all $0\leq \rho(\mathbf{x})< \rho_0$ (see Equation~\eqref{Repulsive_Force}), which shows that the system~\eqref{Single_Integrator} is safe.
\end{proof}
\subsection{Integration of APFs in RCBF-QP Safety Filters}
In this subsection, we aim to integrate the APFs into the RCBF-QP safety filters, thereby formulating a special RCBF-QP safety filter designed for a safe stabilization task of a single-integrator model. To this end, we first design a nominal control law $\mathbf{u}_{\mathrm{nom}}=\mathbf{u}_{\mathrm{CLF}}$ for the dynamical model~\eqref{Single_Integrator}, using an attractive potential field as a CLF.
\begin{Lem}\label{APF_Special_QP}
    Consider the dynamic model~\eqref{Single_Integrator} and set $V(\mathbf{x})=\mathrm{U}_{\mathrm{att}}(\mathbf{x})$, where $\mathrm{U}_{\mathrm{att}}(\mathbf{x})$ is the attractive potential field defined in~\eqref{Attractive}. The control law defined by the following optimization problem satisfies the CLF condition~\eqref{CLF_Condition} and ensures the asymptotic stability of the system~\eqref{Single_Integrator}.
\begin{subequations}\label{APF_CLF_QP}
    \begin{align}
         \mathbf{u}_{\mathrm{CLF}}=&\mathop{\mathrm{argmin}}_{\mathbf{u}\in\mathbb{R}^{m}}\frac{1}{2}\|\mathbf{u}\|^{2}\label{APF_CLF_QP_Cost}\\
    &\,\,\mathrm{s.t.}\,\, \widetilde{a}(\mathbf{x})+\mathbf{b}(\mathbf{x})\mathbf{u}\leq 0, \label{APF_CLF_QP_Con}
    \end{align}
\end{subequations}
where $\widetilde{a}(\mathbf{x})=a(\mathbf{x})+\sigma(\mathbf{x})$, $a(\mathbf{x})=0$, $\sigma(\mathbf{x})=\|\mathbf{b}(\mathbf{x})\|^{2}$, $\mathbf{b}(\mathbf{x})=\mathbf{F}_{\mathrm{att}}(\mathbf{x})^{\top}$. Moreover, the solution to the optimization~\eqref{APF_CLF_QP} is $\mathbf{u}_{\mathrm{CLF}}=-\mathbf{F}_{\mathrm{att}}(\mathbf{x})$.
\end{Lem}
\begin{proof}
    As noticed, the constraint~\eqref{APF_CLF_QP_Con} serves as a sufficient condition for the standard CLF condition~\eqref{CLF_Condition}. Consequently, the solution to~\eqref{APF_CLF_QP} guarantees the satisfaction of the CLF condition, thereby ensuring the asymptotic stability of the closed-loop system~\eqref{Affine_Control_System}. Moreover, the solution to~\eqref{APF_CLF_QP} can be obtained by setting $\mathbf{u}_{\mathrm{nom}}=\mathbf{0}$ in~\eqref{Tightened_CBF_QP} and replacing $\widetilde{c}(\mathbf{x})$ and $\mathbf{d}(\mathbf{x})$ to $\widetilde{a}(\mathbf{x})$ and $\mathbf{b}(\mathbf{x})$, respectively.
\end{proof}
\begin{Rmk}\label{Nominal_CLF_AC}
Note that, in~\eqref{APF_CLF_QP}, any positive definite function $\sigma(\mathbf{x})$ enables the derivation of a nominal controller satisfying the CLF condition~\eqref{CLF_Condition} and ensuring the asymptotic stability of system~\eqref{Single_Integrator}. We specifically choose $\sigma(\mathbf{x})=\|\mathbf{b}(\mathbf{x})\|^2$ in Lemma~\ref{APF_Special_QP}. This choice is by realizing that setting $\sigma(\mathbf{x})=\|\mathbf{b}(\mathbf{x})\|^2$ in~\eqref{APF_CLF_QP} yields a nominal control law $\mathbf{u}_{\mathrm{nom}}=-\mathbf{F}_{\mathrm{att}}(\mathbf{x})$ for a single-integrator model. This is exactly the control law derived from attractive potential fields in~\eqref{Attr_Force}. According to Lemma~\ref{CLF_Att}, the nominal control law $\mathbf{u}_{\mathrm{CLF}}$ guarantees asymptotic stability of the system~\eqref{Single_Integrator}. 
\end{Rmk}
Next, we choose the repulsive potential field to be an RCBF since it ensures the safety guarantees of the system~\eqref{Single_Integrator} as revealed in Lemma~\ref{CBF_Rep}.
\begin{Lem}\label{APF_Special_CBF}
    Consider the single-integrator model~\eqref{Single_Integrator}. We set $\mathbf{u}_{\mathrm{nom}}=\mathbf{u}_{\mathrm{CLF}}=-\mathbf{F}_{\mathrm{att}}(\mathbf{x})$, $B(\mathbf{x})=\mathbf{U}_{\mathrm{rep}}(\mathbf{x})$, $h(\mathbf{x})=\rho(\mathbf{x})$, and $\Gamma(\mathbf{x})=\|\mathbf{F}_{\mathrm{rep}}(\mathbf{x})\|^{2}+\alpha(h(\mathbf{x}))-\mathbf{d}(\mathbf{x})\mathbf{u}_{\mathrm{nom}}$. From the RCBF-QP safety filter in~\eqref{Tightened_CBF_QP}, we obtain the control law $\mathbf{u}_{\mathrm{ACBF}}$, given by:
      \begin{equation}\label{CBF_Special}
        \mathbf{u}_{\mathrm{ACBF}}=\begin{cases}
        -\mathbf{F}_{\mathrm{att}}(\mathbf{x}),&\mathbf{F}_{\mathrm{rep}}(\mathbf{x})=\mathbf{0},\\
        -\mathbf{F}_{\mathrm{att}}(\mathbf{x})-\mathbf{F}_{\mathrm{rep}}(\mathbf{x}),&\mathbf{F}_{\mathrm{rep}}(\mathbf{x})\neq\mathbf{0}.
\end{cases}
    \end{equation}
\end{Lem}
\begin{proof}
Setting $B(\mathbf{x})=\mathbf{U}_{\mathrm{rep}}(\mathbf{x})$ and $h(\mathbf{x})=\rho(\mathbf{x})$, we have $c(\mathbf{x})=-\alpha(h(\mathbf{x}))$ and $\mathbf{d}(\mathbf{x})=\mathbf{F}_{\mathrm{rep}}(\mathbf{x})^{\top}$ for the dynamical model~\eqref{Single_Integrator}, following Definition~\ref{RCBF_Def}. Subsequently, employing the definition of T-RCBF in~\eqref{Tightened_RCBF}, we further determine $\widetilde{c}(\mathbf{x})=\|\mathbf{F}_{\mathrm{rep}}(\mathbf{x})\|^{2}-\mathbf{d}(\mathbf{x})\mathbf{u}_{\mathrm{nom}}$ with $\Gamma(\mathbf{x})=\|\mathbf{F}_{\mathrm{rep}}(\mathbf{x})\|^{2}+\alpha(h(\mathbf{x}))-\mathbf{d}(\mathbf{x})\mathbf{u}_{\mathrm{nom}}$. Then, since $\mathbf{u}_{\mathrm{nom}}=-\mathbf{F}_{\mathrm{att}}(\mathbf{x})$ and $\widetilde{c}(\mathbf{x})=\|\mathbf{F}_{\mathrm{rep}}(\mathbf{x})\|^{2}-\mathbf{d}(\mathbf{x})\mathbf{u}_{\mathrm{nom}}$, a special RCBF-QP safety filter can be defined according to~\eqref{Tightened_CBF_QP}. Specifically, according to Lemma~\ref{Tightened_CBF_QP_Lem}, the solution to the special RCBF-QP safety filter~\eqref{Safety_Filter} gives a control law as in~\eqref{CBF_Special}. 
\end{proof}
\begin{Rmk}
Note that, to establish Lemma~\ref{APF_Special_CBF}, we require an appropriate choice of \(\Gamma(\mathbf{x})\). As mentioned in~\eqref{Tightened_RCBF}, \(\Gamma(\mathbf{x})\) must be positive semidefinite. This condition can be ensured by selecting \(\alpha(h(\mathbf{x}))\) sufficiently large so that the function $\Gamma(\mathbf{x}) = \|\mathbf{F}_{\mathrm{rep}}(\mathbf{x})\|^{2} + \alpha(h(\mathbf{x})) - \mathbf{d}(\mathbf{x}) \mathbf{u}_{\mathrm{nom}}$, remains nonnegative. In particular, requiring $\Gamma(\mathbf{x}) \ge 0$ implies $\alpha(h(\mathbf{x})) \geq \mathbf{d}(\mathbf{x})  \mathbf{u}_{\mathrm{nom}} - \|\mathbf{F}_{\mathrm{rep}}(\mathbf{x})\|^{2}$. Furthermore, following the results in Lemma~\ref{APF_Special_CBF}, we should choose $\alpha(h(\mathbf{x})) \geq -\,\mathbf{F}_{\mathrm{rep}}(\mathbf{x})^\top \mathbf{F}_{\mathrm{att}}(\mathbf{x}) - \|\mathbf{F}_{\mathrm{rep}}(\mathbf{x})\|^{2}$.
\end{Rmk}
\subsection{Equivalence of APFs and RCBF-QP Safety Filters}
In this subsection, we show how APF-designed controllers relate to RCBF-QP safety filters. Specifically, we prove that the APF-designed controller is equivalent to the one defined by the RCBF-QP safety filters presented in Lemma~\ref{APF_Special_CBF}.
\begin{Thm}\label{Main_Theorem}
    Consider the single-integrator model~\eqref{Single_Integrator}. We define the Lyapunov function as $V(\mathbf{x})=\mathrm{U}_{\mathrm{att}}(\mathbf{x})$, $\sigma(\mathbf{x})=\|\mathbf{b}(\mathbf{x})\|^{2}$, and use~\eqref{APF_CLF_QP} to design the nominal control law $\mathbf{u}_{\mathrm{nom}}=\mathbf{u}_{\mathrm{CLF}}$. Moreover, we set $B(\mathbf{x})=\mathbf{U}_{\mathrm{rep}}(\mathbf{x})$, $h(\mathbf{x})=\rho(\mathbf{x})$, and $\Gamma(\mathbf{x})=\|\mathbf{d}(\mathbf{x})\|^{2}+\alpha(h(\mathbf{x}))-\mathbf{d}(\mathbf{x})\mathbf{u}_{\mathrm{nom}}$. The APF-designed controller~\eqref{APF_Force} is then equivalent to the controller obtained by the RCBF-QP safety filter given in~\eqref{CBF_Special}.
\end{Thm}
\begin{proof}
Based on Lemma~\ref{APF_Special_QP}, we first derive the nominal control law \(\mathbf{u}_{\mathrm{nom}} = \mathbf{u}_{\mathrm{CLF}}\). Then, applying Lemma~\ref{APF_Special_CBF}, we will obtain the control law 
\(\mathbf{u}_{\mathrm{ACBF}}\) from the RCBF-QP safety filter, as defined in~\eqref{CBF_Special}. Next, we compare the APF-designed controller~\eqref{APF_Force} with the one 
obtained from the RCBF-QP safety filter~\eqref{CBF_Special}. By noticing that in~\eqref{CBF_Special}, the repulsive force 
\(\mathbf{F}_{\mathrm{rep}}(\mathbf{x})\) is zero if and only if 
\(\rho(\mathbf{x}) \geq \rho_0\), and nonzero if and only if 
\(\rho(\mathbf{x}) < \rho_0\). Hence, the conditions
\(\mathbf{F}_{\mathrm{rep}}(\mathbf{x}) = \mathbf{0}\) and 
\(\mathbf{F}_{\mathrm{rep}}(\mathbf{x}) \neq \mathbf{0}\) in~\eqref{CBF_Special}
can be mutually exchanged with \(\rho(\mathbf{x}) \geq \rho_0\) and 
\(\rho(\mathbf{x}) < \rho_0\) in~\eqref{APF_Force}, respectively. Therefore, it shows that the control law of the APF-based method~\eqref{APF_Force} and the RCBF-QP safety filter~\eqref{CBF_Special} give the same controller, which demonstrates their equivalence.
\end{proof}
As stated in Theorem~\ref{Main_Theorem}, we establish the equivalence between APF-based controllers and RCBF-QP safety filters by integrating APFs into the RCBF-QP framework. A critical factor in this equivalence is the choice of $\sigma(\mathbf{x})$ and $\Gamma(\mathbf{x})$. However, as shown in~\eqref{Tightened_CLF_Condition} and~\eqref{Tightened_RCBF}, the only requirement for \(\sigma(\mathbf{x})\) is that it be positive definite, while \(\Gamma(\mathbf{x})\) must be positive semidefinite. In Theorem~\ref{Main_Theorem}, we chose specific forms for \(\sigma(\mathbf{x})\) and \(\Gamma(\mathbf{x})\) due to the objective to establish the equivalence between the APF-designed controller in~\eqref{APF_Force} and the RCBF-QP safety filter in~\eqref{CBF_Special}. If we relax this requirement and do not require specific forms on \(\sigma(\mathbf{x})\) and \(\Gamma(\mathbf{x})\), we should obtain a more generalized APF-based controller, or equivalently a more general RCBF-QP safety filter-based controller.
\subsection{A Generalized APF-based Controller}
Now, we aim to derive a generalized APF-based controller by extending the results in Theorem~\ref{Main_Theorem}, without restricting the selections of \(\sigma(\mathbf{x})\) and \(\Gamma(\mathbf{x})\).
\begin{Thm}\label{Main_Theorem_2}
    Consider the single-integrator model~\eqref{Single_Integrator}. We define the Lyapunov function as $V(\mathbf{x})=\mathrm{U}_{\mathrm{att}}(\mathbf{x})$, $\sigma(\mathbf{x})$ is a positive definite function, and use~\eqref{APF_CLF_QP} to design the nominal control law $\mathbf{u}_{\mathrm{nom}}$. Moreover, we set $B(\mathbf{x})=\mathbf{U}_{\mathrm{rep}}(\mathbf{x})$, $h(\mathbf{x})=\rho(\mathbf{x})$, and $\Gamma(\mathbf{x})$ is a positive semidefinite function. The generalized APF-based controller derived using the RCBF-QP safety filter presented in~\eqref{Tightened_CBF_QP} is given below.
\begin{equation}\label{eq:APF-SF-SI}
    \mathbf{u}_{{\scriptscriptstyle \textnormal{GAPF}}} =
    \begin{cases}
        G_{\textnormal{att}}(\mathbf{x})\mathbf{F}_{\textnormal{att}}(\mathbf{x}), 
        & G_{\textnormal{rep}}(\mathbf{x}) \leq 0, \\
        G_{\textnormal{att}}(\mathbf{x})\mathbf{F}_{\textnormal{att}}(\mathbf{x}) 
         + G_{\textnormal{rep}}(\mathbf{x})\mathbf{F}_{\textnormal{rep}}(\mathbf{x}),
        & G_{\textnormal{rep}}(\mathbf{x}) > 0,
    \end{cases}
\end{equation}
where $G_{\textnormal{att}}(\mathbf{x})$ and $G_{\textnormal{rep}}(\mathbf{x})$ are given by
\begin{equation*}\label{eq:APF-SF-weights}
    \begin{aligned}
        G_{\textnormal{att}}(\mathbf{x}) &=- \frac{\sigma(\mathbf{x})}{||\mathbf{F}_{\textnormal{att}}(\mathbf{x})||^2},\\
        G_{\textnormal{rep}}(\mathbf{x}) &= -\frac{\Gamma(\mathbf{x}) - \alpha(h(\mathbf{x})) + G_{\textnormal{att}}(\mathbf{x})\mathbf{F}_{\textnormal{rep}}(\mathbf{x})^\top\mathbf{F}_{\textnormal{att}}(\mathbf{x})}{||\mathbf{F}_{\textnormal{rep}}(\mathbf{x})||^2}.
    \end{aligned}
\end{equation*}
\end{Thm}
\begin{proof}
Firstly, for the dynamical system~\eqref{Single_Integrator}, we can easily have $\widetilde{a}(\mathbf{x})=a(\mathbf{x})+\sigma(\mathbf{x})$, $a(\mathbf{x})=0$, and $\mathbf{b}(\mathbf{x})=\mathbf{F}_{\mathrm{att}}(\mathbf{x})^{\top}$ according to~\eqref{CLF_Condition} and~\eqref{Tightened_CLF_Condition}. Then, based on the results presented in Lemma~\ref{Tightened_CBF_QP_Lem}, we obtain that the solution to~\eqref{APF_CLF_QP} is given by $\mathbf{u}_{\mathrm{nom}}=G_{\textnormal{att}}(\mathbf{x})\mathbf{F}_{\textnormal{att}}(\mathbf{x})$. Next, with $\widetilde{c}(\mathbf{x})=c(\mathbf{x})+\Gamma(\mathbf{x})$, $c(\mathbf{x})=-\alpha(h(\mathbf{x}))$, $\mathbf{d}(\mathbf{x})=\mathbf{F}_{\mathrm{rep}}(\mathbf{x})^{\top}$, and $\Gamma(\mathbf{x})$ is positive semidefinite, substituting $\mathbf{u}_{\mathrm{nom}}$ into~\eqref{Tightened_CBF_QP} gives the solution in~\eqref{eq:APF-SF-SI} by using Lemma~\ref{Tightened_CBF_QP_Lem}.
\end{proof}
\begin{Rmk}
    Note that, by choosing $\sigma(\mathbf{x})=\|\mathbf{b}(\mathbf{x})\|^{2}$ and $\Gamma(\mathbf{x})=\|\mathbf{d}(\mathbf{x})\|^{2}+\alpha(h(\mathbf{x}))-\mathbf{d}(\mathbf{x})\mathbf{u}_{\mathrm{nom}}$, Theorem~\ref{Main_Theorem_2} will provide us with the APF-based controller given in~\eqref{APF_Force}, or equivalently, the RCBF-QP safety filter presented in~\eqref{CBF_Special}. This demonstrates that the control law in~\eqref{eq:APF-SF-SI} generalizes both the APF-based controller~\eqref{APF_Force} and the RCBF-QP safety filter~\eqref{CBF_Special}. As one can expect, any combination of $\sigma(\mathbf{x})$ and $\Gamma(\mathbf{x})$ will give us a new controller.
\end{Rmk}
\section{Simulations}
In this section, we show the connections between APF-based controllers and RCBF-QP safety filters using a simple example of collision avoidance. We show that the equivalence between these two approaches holds only when selecting $\sigma(\mathbf{x})=\|\mathbf{b}(\mathbf{x})\|^{2}$ and $\Gamma(\mathbf{x})=\|\mathbf{d}(\mathbf{x})\|^{2}+\alpha(h(\mathbf{x}))-\mathbf{d}(\mathbf{x})\mathbf{u}_{\mathrm{nom}}$, as established by Theorem~\ref{Main_Theorem}.

Consider the single-integrator system~\eqref{Single_Integrator}, where the state $\mathbf{x}$ represents the position of an agent. The goal is to reach the desired position $\mathbf{x}_{\mathrm{goal}} = [7, 3.2]^{\top}$. In addition, the centers of the three obstacles are $\mathbf{x}_{\mathrm{O_1}}=[-0.4, 1.5]^{\top}$, $\mathbf{x}_{\mathrm{O_2}}=[2.0,3.3]^{\top}$, and $\mathbf{x}_{\mathrm{O_3}}=[4.5, 2.5]^{\top}$, and their radius is set to $r=0.5$. The basic settings for defining the attractive potential in~\eqref{Attractive} and the repulsive potential in~\eqref{Repulsive} are given as follows: $K_{\mathrm{att}}=K_{\mathrm{rep}}=1$, $\alpha(h(\mathbf{x}))=h(\mathbf{x})$, and $\rho_{0}=0.2$. Following this, we utilize the attractive potential field defined in~\eqref{Attractive} and the repulsive potential field expressed in~\eqref{Repulsive} to serve as CLF and RCBF, respectively. Moreover, according to the results of Theorem~\ref{Main_Theorem_2}, we explore several
generalized APF-based controllers by fixing $\sigma(\mathbf{x})=\|\mathbf{b}(\mathbf{x})\|^{2}$ and discussing
$\Gamma(\mathbf{x})$ in three cases. Specifically, we set $\Gamma_{1}(\mathbf{x}) = 0$,
$\Gamma_{2}(\mathbf{x})= 8\|\mathbf{d}(\mathbf{x})\|^2 + \alpha\bigl(h(\mathbf{x})\bigr)- \mathbf{d}(\mathbf{x})^\top \mathbf{u}_{\mathrm{nom}}$, and $\Gamma_{3}(\mathbf{x}) = \|\mathbf{d}(\mathbf{x})\|^2 + \alpha\bigl(h(\mathbf{x})\bigr)- \mathbf{d}(\mathbf{x})^\top \mathbf{u}_{\mathrm{nom}}$. In the first case, $\Gamma_{1}(\mathbf{x})=0$ indicates that the RCBF in Definition~\ref{RCBF_Def} is used to obtain a QP-based controller instead of the T-RCBF given~\eqref{Tightened_CBF_QP}. In the second case, we emphasize that an arbitrary choice of $\Gamma(\mathbf{x})=\Gamma_{2}(\mathbf{x})$ can provide a new controller. The third case is associated with the RCBF-QP safety filter in~\eqref{CBF_Special}. Furthermore, to show the equivalence between the APF-designed controller and the one obtained by RCBF-QP safety filter, the APF-designed controller~\eqref{APF_Force} is also provided in the simulations.

\begin{figure}[tp]
 \centering
    \makebox[0pt]{%
    \includegraphics[width=0.49\textwidth]{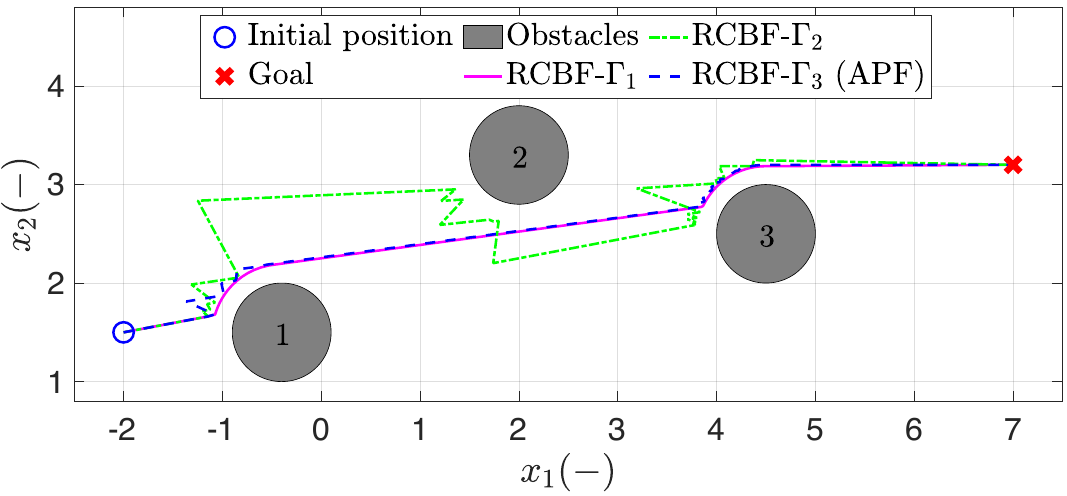}}
    \caption{Different
generalized APF-based controllers with different
$\Gamma(\mathbf{x})$ in three cases.}
    \label{APF_AC_Controller}
\end{figure}

As shown in Fig.~\ref{APF_AC_Controller}, all three controllers parameterized by different values of $\Gamma$ can successfully ensure the safety of the single-integrator system~\eqref{Single_Integrator}. However, they have different performances. Specifically, the new controller with $\Gamma(\mathbf{x})=\Gamma_{2}(\mathbf{x})$ and the APF-designed controller with $\Gamma(\mathbf{x})=\Gamma_{3}(\mathbf{x})$ tend to produce obvious oscillations in the navigation trajectory. In contrast, selecting a suitable $\Gamma$ ($\Gamma=\Gamma_{1} = 0$ in the simulation) results in much smoother trajectories under the RCBF-QP safety filter. This finding is consistent with the comparative observations in~\cite{Comparative_Study}, which further suggests that RCBF-QP safety filters outperform APF methods in terms of obtaining smooth trajectories.
\section{Conclusions}
This paper demonstrates the equivalence between safety controllers designed using APFs and RCBF-QP safety filters. By introducing T-CLFs and T-RCBFs, we explicitly bridge the gap between APF and RCBF-QP frameworks. Specifically, an attractive potential field serves as a T-CLF for nominal controller design, while a repulsive potential field is employed as a T-RCBF within the RCBF-QP formulation. Our findings demonstrate that with suitably chosen auxiliary functions in T-CLFs and T-RCBFs, APF-based and RCBF-QP controllers are equivalent. Furthermore, we extend the two methodologies to more general scenarios without restricting the selection of auxiliary functions. Finally, a collision avoidance example is presented, which shows the connection and equivalence between the two methods. 
\bibliographystyle{IEEEtran}
\bibliography{APF_Uni.bib}
\end{document}